\documentclass[10pt, conference]{IEEEtran}
\usepackage[utf8]{inputenc}
\usepackage{graphicx}
\usepackage{caption}
\usepackage{subcaption}
\usepackage{amsmath}
\usepackage{multirow}
\usepackage[table]{xcolor}
\usepackage{pifont}
\newcommand{\cmark}{\ding{51}}
\newcommand{\xmark}{\ding{55}}
\usepackage{array}
\newcolumntype{L}[1]{>{\raggedright\let\newline\\\arraybackslash\hspace{0pt}}m{#1}}
\newcolumntype{C}[1]{>{\centering\let\newline\\\arraybackslash\hspace{0pt}}m{#1}}
\newcolumntype{R}[1]{>{\raggedleft\let\newline\\\arraybackslash\hspace{0pt}}m{#1}}
\usepackage{makecell}
\usepackage{stfloats}

\usepackage{tikz}
\usepackage{textcomp}
\usepackage{hyperref}
\newcommand\copyrighttext{%
  \footnotesize \textcopyright 2023 IEEE. Personal use of this material is permitted.
  Permission from IEEE must be obtained for all other uses, in any current or future 
  media, including reprinting/republishing this material for advertising or promotional 
  purposes, creating new collective works, for resale or redistribution to servers or 
  lists, or reuse of any copyrighted component of this work in other works. 
   DOI: \href{<https://ieeexplore.ieee.org/document/10202778>}{10.1109/PowerTech55446.2023.10202778}
  }
\newcommand\copyrightnotice{%
\begin{tikzpicture}[remember picture,overlay]
\node[anchor=south,yshift=10pt] at (current page.south) {\fbox{\parbox{\dimexpr\textwidth-\fboxsep-\fboxrule\relax}{\copyrighttext}}};
\end{tikzpicture}%
}

\title{Vehicle-to-grid plug-in forecasting for participation in ancillary services markets}
\author{\IEEEauthorblockN{Jemima Graham}
\IEEEauthorblockA{ 
\textit{Imperial College London}\\
London, United Kingdom \\
jemima.graham16@imperial.ac.uk}
\and
\IEEEauthorblockN{Fei Teng}
\IEEEauthorblockA{ 
\textit{Imperial College London}\\
London, United Kingdom\\
f.teng@imperial.ac.uk}}

\date{October 2022}

\begin{document}

\maketitle

\copyrightnotice

\begin{abstract}
    Electric vehicle (EV) charge points (CPs) can be used by aggregators to provide frequency response (FR) services. Aggregators must have day-ahead half-hourly forecasts of minimum aggregate vehicle-to-grid (V2G) plug-in to produce meaningful bids for the day-ahead ancillary services market. However, there is a lack of understanding on what features should be considered and how complex the forecasting model should be. This paper explores the dependency of aggregate V2G plug-in on historic plug-in levels, calendar variables, and weather conditions. These investigations are used to develop three day-ahead forecasts of minimum aggregate V2G plug-in during 30-minute window. A neural network that considers previous V2G plug-in values the day before, three days before, and seven days before, in addition to day of the week, month, and hour, is found to be the most accurate.
\end{abstract}


\section{Introduction}

As inverter-based resource (IBR) penetration in power systems increases, power system inertia declines \cite{Heylen2021, Milano2018}. 
This makes frequency response (FR) increasingly important to maintain the system security. 
Alternative frequency services are now essential to ensure that the power system frequency remains within acceptable limits at all times \cite{Chu2020}. This paper focuses on primary frequency response (PFR) services which respond to a disturbance in less than 10 seconds, providing up to 20 seconds of support \cite{NGESO_FFR_website}.

One source of FR that has long been discussed is electric vehicles (EVs) with vehicle-to-grid (V2G) capabilities \cite{Kempton2005, Kempton2005a}. This is an attractive solution as the number of EVs has been exploding with 23.2 million EVs expected by 2032 \cite{Wills2020}. Moreover, the value of using V2G for FR has been confirmed in studies that examine the reduction in frequency deviations \cite{Meng2015} and subsequent cost savings that can be achieved using this technology \cite{OMalley2022, OMalley2020}. This paper focuses on bidirectional V2Gs that can provide up-regulation PFR by supplying a net power injection to the grid.

It is expected that V2G market participation and dispatch will be managed through an aggregator because current market rules expect a minimum of 1 MW capacity and V2G chargers have a maximum capacity of 15 kW \cite{Kempton2005a}. Some business cases for who could take on the role of an aggregator are: an EV fleet operator; an electric utility company; or an independent company such as automobile manufacturer or a distributed generation manager \cite{Kempton2005a}. Here, we consider an electric utility company that owns a number of charge points (CPs) across the UK. This presents unique challenges as the aggregator will not have a schedule for EV plug-in as an EV fleet operator might have.

The amount of up-regulation that V2G can provide depends on: (i) whether it is plugged in (so that it can discharge to the grid); and (ii) whether it is charging (so that it can stop charging) \cite{Kempton2005a}. Day-ahead forecasts of aggregate V2G availability are necessary to allow the aggregator to produce day-ahead bids for the ancillary services markets \cite{Lopes2011, Han2019}. It is important to have separate forecasts of aggregate V2G plug-in and aggregate V2G charging due to the different ways of providing up-regulation using V2G. Here, it is assumed that there is uniform discharging capability across CPs and that PFR has minimal impact on the energy stored in the batteries. Half-hourly forecasts are necessary here to match the bidding frequency in the regulation market.

V2G availability and EV user behaviour are forecast in the literature to inform aggregator decision-making independently or as an input to a scheduling optimization model. Even though the focus in this study is aggregator information, the forecasting models developed as part of scheduling processes will also be discussed. These works are outlined in Table \ref{tab:existing_work}.

\begin{table*}[ht!]
    \centering
    \caption{Comparison between existing EV forecasts and the aggregate EV plug-in forecast models proposed here.}
    \begin{tabular}{|L{1cm}|C{1.2cm}|L{2cm}|C{1.1cm}|C{1.1cm}|C{1.15cm}|C{1.1cm}|C{1.1cm}|L{2.5cm}|C{0.5cm}|C{0.5cm}|}
    \hline
      \multirow{2}{1cm}{\centering \textbf{Study}} & \multirow{2}{1.2cm}{\centering \textbf{Aggregate}} & 
      \multirow{2}{2cm}{\centering \textbf{Target}} & \multirow{2}{1.1cm}{\centering \textbf{Include weekend}} & 
      \multicolumn{4}{|c|}{\textbf{Features}} & \multirow{2}{2.5cm}{\centering \textbf{Model}} & \multicolumn{2}{|c|}{\centering \textbf{Dataset}} \\
    \cline{5-8} \cline{10-11}
     &  &  &  & \textbf{Historic} & \textbf{Calendar} & \textbf{Weather} & \textbf{Other} & & \textbf{CPs} & \textbf{EVs} \\
     \Xhline{5\arrayrulewidth}
     \cite{AmaraOuali2022} & \cellcolor{green!25}\cmark & Occupancy and charging load & \cellcolor{red!25}\xmark & \cellcolor{green!25}\cmark & \cellcolor{green!25}\cmark & \cellcolor{red!25}\xmark & \cellcolor{red!25}\xmark & Generalized additive model, random forest, SARIMA & \cellcolor{green!25}\cmark & \cellcolor{red!25}\xmark \\
     \hline
     \cite{Dominguez2021} & \cellcolor{red!25}\xmark & Trip start and end location, and trip distance & \cellcolor{green!25}\cmark & \cellcolor{red!25}\xmark & \cellcolor{green!25}\cmark & \cellcolor{red!25}\xmark & \cellcolor{red!25}\xmark & LightGBM & \cellcolor{red!25}\xmark & \cellcolor{green!25}\cmark \\
     \hline
     \cite{Li2021} & \cellcolor{green!25}\cmark  & Scheduable energy capacity & \cellcolor{red!25}\xmark & \cellcolor{green!25}\cmark & \cellcolor{red!25}\xmark & \cellcolor{red!25}\xmark & \cellcolor{red!25}\xmark & Offline (day-ahead)/ rolling (hour-ahead) LSTM & \cellcolor{red!25}\xmark & \cellcolor{green!25}\cmark \\
     \hline
     \cite{Perry2021} & \cellcolor{red!25}\xmark & Hour-ahead energy demand & \cellcolor{green!25}\cmark &  \cellcolor{green!25}\cmark & \cellcolor{red!25}\xmark & \cellcolor{red!25}\xmark & \cellcolor{red!25}\xmark & K-Means clustering, LSTM using federated learning & \cellcolor{green!25}\cmark & \cellcolor{red!25}\xmark \\
     \hline
     \cite{Giordano2020} & \cellcolor{red!25} \xmark & Arrival/departure times, energy for next trip & \cellcolor{green!25}\cmark & \cellcolor{green!25}\cmark & \cellcolor{green!25}\cmark & \cellcolor{red!25}\xmark & \cellcolor{red!25}\xmark & K-medoid clustering of statistical modes & \cellcolor{red!25}\xmark & \cellcolor{green!25}\cmark \\
     \hline
     \cite{Huber2020} & \cellcolor{red!25}\xmark & Parking duration, trip distance & \cellcolor{green!25}\cmark & \cellcolor{green!25}\cmark & \cellcolor{green!25}\cmark & \cellcolor{red!25}\xmark & \cellcolor{green!25}User type & Quantile regression, MLP, kernel density estimation & \cellcolor{red!25}\xmark & \cellcolor{green!25}\cmark \\
     \hline
     \cite{Jahangir2019} & \cellcolor{red!25}\xmark & Arrival time, departure time, and trip distance & \cellcolor{green!25}\cmark & \cellcolor{green!25}\cmark & \cellcolor{red!25}\xmark & \cellcolor{red!25}\xmark & \cellcolor{red!25}\xmark & Neural network with Rough structure & \cellcolor{red!25}\xmark & \cellcolor{green!25}\cmark \\
     \hline
     \cite{Saputra2019} & \cellcolor{red!25}\xmark & Energy demand at a given time & \cellcolor{green!25}\cmark & \cellcolor{red!25}\xmark & \cellcolor{green!25}\cmark & \cellcolor{red!25}\xmark & \cellcolor{green!25}Location & K-Means clustering, neural network using federated learning & \cellcolor{green!25}\cmark & \cellcolor{red!25}\xmark \\
     \hline
     \cite{Bessa2013, Bessa2013a} & \cellcolor{green!25}\cmark & Charging distribution, max power available & \cellcolor{green!25}\cmark & \cellcolor{green!25}\cmark & \cellcolor{green!25}\cmark & \cellcolor{red!25}\xmark & \cellcolor{red!25}\xmark & Generalized additive model & \cellcolor{red!25}\xmark & \cellcolor{green!25}\cmark\\
     \Xhline{5\arrayrulewidth}
     \textbf{Our models} & \cellcolor{green!25}\cmark & \textbf{Half-hourly minimum of aggregate EV plug-in} & \cellcolor{green!25}\cmark & \cellcolor{green!25}\cmark & \cellcolor{green!25}\cmark & \cellcolor{orange!25}\textbf{Investigated} & \cellcolor{red!25}\xmark & \textbf{Persistence model, generalized linear model, neural network} & \cellcolor{green!25}\cmark & \cellcolor{red!25}\xmark \\
     \hline
    \end{tabular}
    \label{tab:existing_work}
\end{table*}

There is little work in the field of aggregate EV plug-in forecasting. Only two studies in Table \ref{tab:existing_work} attempt to directly predict the aggregate availability of V2G \cite{AmaraOuali2022, Li2021}. While one of these works considers an aggregator who controls a fleet of EVs \cite{Li2021}, the other considers an aggregator only has access to CP data \cite{AmaraOuali2022}. However, the latter model developed in \cite{AmaraOuali2022} is a minutely forecast model that excludes weekends. Not only this, it has limited utility to aggregators who are risk-averse as it is a deterministic forecast. In our study, the forecast models, which capture behaviour on both weekdays and weekends, aim to forecast the minimum aggregate plug-in. This allows aggregators to manage their risk and avoid significant penalty fees for failing to deliver any contracted services \cite{Han2019, Clairand2019, Clairand2018}. 

Generally, it is important to make a distinction between the studies that use EV fleet data and the studies that use CP data because EV fleet operators have access to information that would not be accessible to CP operators such as personal charging preferences or trip schedule \cite{Huber2020, Saputra2019}. Additionally, none of the existing work in Table \ref{tab:existing_work} considered weather variables as model inputs. Our consideration of weather data is inspired by Gautam et al. who consider temperature, humidity, wind speed, rainfall, and dewpoint temperature to forecast grid load as an input to an EV scheduling algorithm \cite{Gautam2019}.

Altogether, this study aims to develop a first-of-its-kind, day-ahead forecast of minimum aggregate V2G plug-in during a 30-minute window using CP data to aid aggregator decision-making for FR. It examines the dependence of aggregate V2G plug-in on: historic behaviour \cite{Li2021, Bessa2013, Giordano2020, Gautam2019}; calendar variables \cite{Huber2020, Gautam2019}; and weather data \cite{Gautam2019}. Additionally, three forecast models are developed and validated on a UK dataset of non-domestic EV charge point data provided by the UK Department of Transport (DoT). The forecast models discussed here are: a persistence model; a generalized linear model; and a neural network. 

The structure of this paper is as follows: Section \ref{sec:case-study} introduces the DoT case study used for feature investigation and model validation; Section \ref{sec:investigation} examines the features that are useful to consider when developing an aggregate V2G plug-in forecast; Section \ref{sec:forecast} explores the day-ahead aggregate V2G plug-in forecasts developed here; and Section \ref{sec:conclusion} discusses directions for future work.

\section{Data collection, processing, and analysis}
\label{sec:case-study}

This work is validated on a dataset of EV plug-in events at public sector CPs during 2017 provided by the UK Department of Transport (DoT). We assume that the EV user behaviour in this dataset is representative of the behaviour of V2G users. Additionally, we consider a public sector dataset as we assume that domestic users will have a good idea of when they will plug-in and could send a potential plug-in schedule to an aggregator at least a day in advance. This dataset is a list of charging events. It had the following features: charging event ID, charge point ID, connector (as each charge point could have more than one connector), start date, start time, end date, end time, energy, public sector name, and plug-in duration. Charging events with a plug-in duration of more than a week are not considered in this study as we are only interested in forecasting the availability of charge points that are actively being used. This excluded 207 charging events, leaving a total of 103119 charging events. Additionally, some charge points that only have one connector did not specify a connector, so these NaN values were filled with ones.

The charge points in the public sector dataset are owned by 35 organizations. Milton Keynes Council has the most charge points with a total of 97, while Bristol City Council and South Tyneside Borough Council have only one charge point each. It is unknown how many charge points the Department for Regional Development Northern Ireland has as they have not supplied charge point IDs; however, as we are interested in aggregate plug-ins and not individual charge points, this data can still be included.

To obtain a time-series of aggregate EV plug-in for all the public sector charge points in this dataset, the start dates, start times, end dates, and end times were used to create a minutely time series for each charging event where ones indicated a plug-in at that time step and zeros denoted no plug-in. These individual charging event time-series were then summed to get an aggregate EV plug-in time-series. After that, the minimum within each half hour was taken, resampling the minutely time-series to a half-hourly time-series.

The first week of data was excluded as we do not know how many EVs were plugged-in at the beginning of 2017. Additionally, public holidays in the UK are excluded as anomalous EV user behaviour occurs on these days and one year of data is not sufficient to capture this behaviour. Finally, the last two weeks of data were excluded as the anomalous behaviour around Christmas exists both before and after Christmas, likely because many people take extended holidays at that time of the year. 

As only one year of data is available, it is difficult for any models to fully capture the annual behaviour if the training and test sets are divided up in chronological order. Instead, the time steps are shuffled after feature creation, with 80\% randomly sampled for the training set, 10 \% for the validation set, and 10 \% for the test set. This does not lead to data leakage as the aggregate EV plug-in time-series is stationary as confirmed by an Augmented Dickey Fuller Test.

\section{Investigation of electric vehicle plug-in characteristics}
\label{sec:investigation}

\begin{figure}
    \centering
    \includegraphics[width=0.43\textwidth]{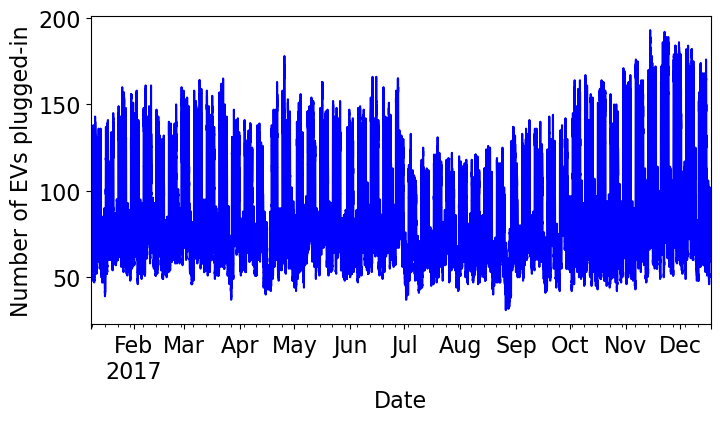}
    \caption{Aggregate EV plug-in calculated from the UK Department of Transport public sector fasts dataset which covers EV plug-in events during 2017.}
    \label{fig:aggregate_ev_plug-in}
\end{figure}

Aggregate EV plug-in is a stochastic time-series as shown in Figure \ref{fig:aggregate_ev_plug-in}. This is because EV user behaviour is unpredictable and varies without a clearly discernible pattern. As a result, aggregate EV plug-in is difficult to predict accurately as current plug-in numbers are not necessarily dependent on previous plug-in numbers.

\begin{figure}[ht!]
    \centering
    \includegraphics[width=0.48\textwidth]{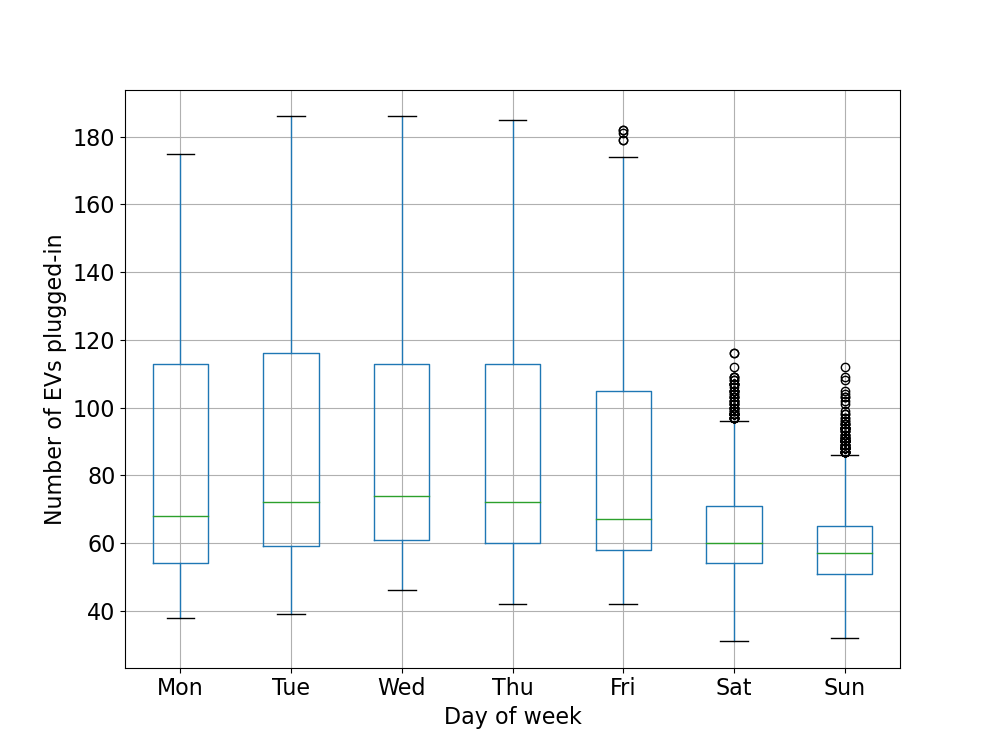}
    \caption{EV plug-in by day of the week. Weekdays have similar medians and interquartile ranges compared to weekends.}
    \label{fig:plug-in_by_day}
\end{figure}

Overall, aggregate EV plug-in characteristics depend on the day of the week as shown in Figure \ref{fig:plug-in_by_day}. Monday through Friday show much higher EV plug-in rates compared to Saturday and Sunday. Also, there is more variability on weekdays than weekends. This suggests that if previous values of EV plug-in are used as a guide in forecasts, it is not sufficient to use the day before in all cases. Our persistence forecast takes these variations into account, thus providing a strong benchmark. 

\begin{figure}[ht!]
    \centering
    \includegraphics[width=0.48\textwidth]{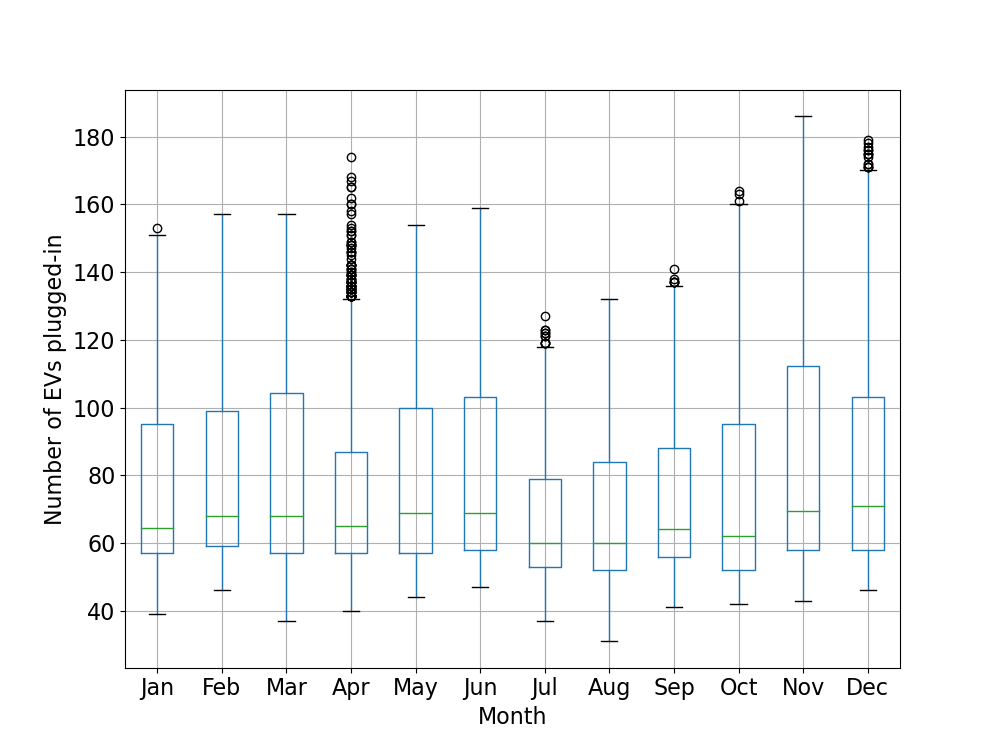}
    \caption{EV plug-in by month. The median number of EV plug-ins each month is similar; however, both the ranges and the interquartile ranges vary vastly from month to month with no clear seasonality.}
    \label{fig:plug-in_by_month}
\end{figure}

Another feature worth considering is the month: while the median number of EV plug-ins is similar each month as shown in Figure \ref{fig:plug-in_by_month}, the range of aggregate EV plug-ins varies greatly from month-to-month. Moreover, these ranges exhibit no clear seasonality. This suggests that including the month as a feature would be beneficial to better capture the range of EV user behaviour depending on the time of year. This variation could be linked to EV user behaviour. For example, there may be a tendency to go on holiday at certain times of the year as seen in the decreased median and interquartile ranges in July and August: peak season for holidaymakers in the UK.

\begin{figure}[ht!]
    \centering
    \includegraphics[width=0.45\textwidth]{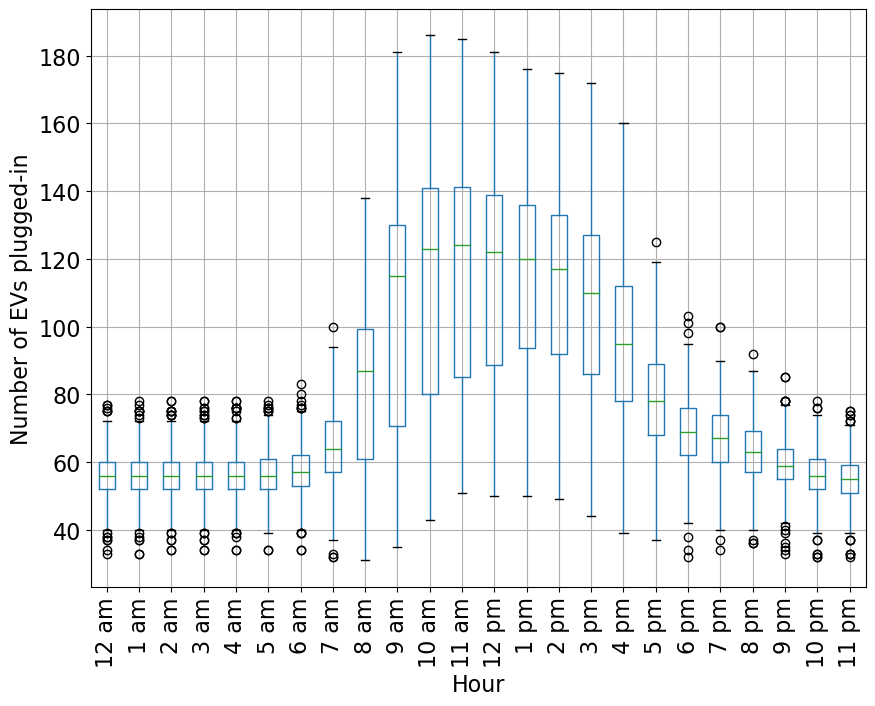}
    \caption{EV plug-in by hour. The median is lower and interquartile ranges are reduced overnight (6 pm - 6 am); there is high variability and increased median plug-in rates during the day.}
    \label{fig:plug-in_by_hour}
\end{figure}

In addition to day of the week and month, EV plug-in also varies by hour as depicted in Figure \ref{fig:plug-in_by_hour}. It can be seen that there is less variation overnight between 6 pm and 6 am. There is also a lower median number of plug-ins during this time. This may be because public charge points are more likely to be accessed during the day when people are at work or visiting establishments that are open during working hours. This hourly variability must also be captured in any forecasting models.

A final area of investigation is the dependence of EV plug-in on weather data: no dependence was found. Due to a lack of day-ahead weather forecasts of a sufficiently fine temporal resolution, MIDAS open-source hourly weather observations from numerous weather stations across the UK were averaged and interpolated to half-hourly values. This weather data covered relative humidity, dewpoint temperature, air temperature, wind speed, and amount of precipitation. Pearson's correlation coefficients to evaluate any linear dependencies and Spearman's correlation coefficients to evaluate any non-linear dependencies were calculated between each of the weather variables and the aggregate EV plug-in. Both the Pearson's and Spearman's correlation coefficients were negligible. Consequently, weather data was not considered during EV plug-in forecast development under this study.

\section{Aggregate V2G plug-in forecast development}
\label{sec:forecast}

In this section, three day-ahead forecast models of increasing complexity are developed to predict the minimum aggregate V2G plug-in in a 30 minute period.

\subsection{Persistence model (PM)}

To capture the behaviour exhibited in Figure \ref{fig:plug-in_by_day}, the persistence forecast uses different historical values depending on which day of the week it is:
\begin{equation}
\hat{y}_{t+48, d} =
\begin{cases}
    y_{t, d-1} & \text{if $d \in \{1, 2, 3, 4, 6\}$} \\
    y_{t-96, 4} & \text{if $d = 0$} \\
    y_{t-288, 5}  & \text{if $d = 5$} \\
\end{cases}
\end{equation}
where $y$ denotes number of V2Gs plugged-in, $d \in \{0, 1, 2, 3, 4, 5, 6\}$ is equivalent to  \{Monday, Tuesday, Wednesday, Thursday, Friday, Saturday, Sunday\} respectively. The subscript $t$ denotes the time step. 
This model is used as a benchmark for the generalized linear model (GLM) and the neural networks (NNs) discussed in the remaining portion of this section.

\subsection{Generalized linear model (GLM)}

A generalized linear model (GLM) is employed to better capture the periodic relationship between V2G plug-in and historic values of V2G plug-in. This forecast model has the following form:

\begin{equation}
    \hat{y}_{t+48} = \alpha_{0, d} y_{t} + \alpha_{-96, d} y_{t-96} + \alpha_{-288, d} y_{t-288}
\end{equation}
where $\alpha_{0, d}$, $\alpha_{-96, d}$, and $\alpha_{-288, d}$ are coefficients of the historic values of V2G plug-in one day before, three days before, and seven days before respectively. The subscript $d$ indicates that there is a different coefficient for each day of the week.

\begin{figure*}
    \centering
\includegraphics[width=0.97\textwidth]{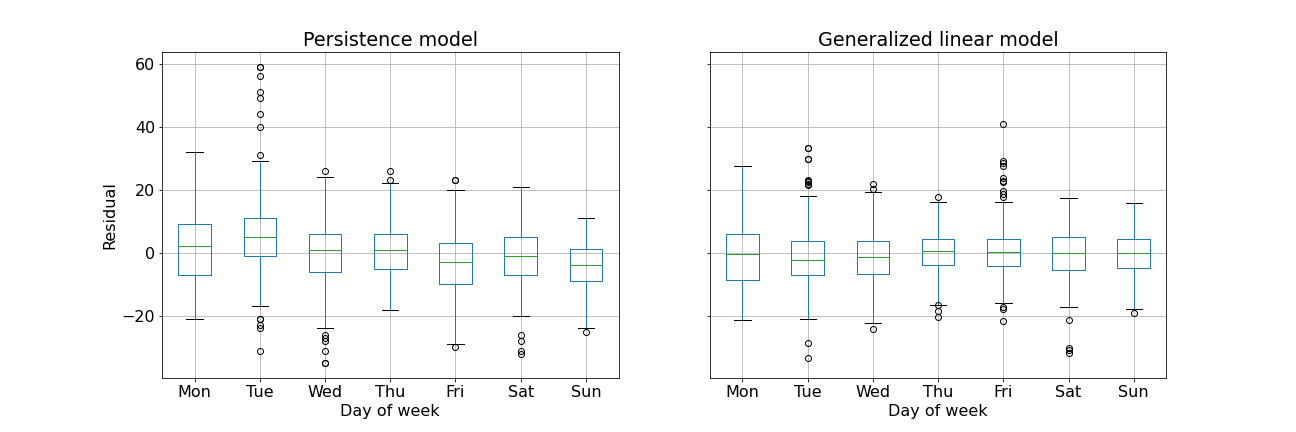}
    \caption{Residuals of the test set for the persistence forecast model (left) and GLM (right). The medians produced by the GLM forecast model are closer to zero than those produced by the persistence forecast. Also, the variability of the residuals is reduced compared to those produced by the persistence forecast.}
    \label{fig:test_resid_by_day}
\end{figure*}

Figure \ref{fig:test_resid_by_day} highlights that using the GLM instead of the persistence model leads to less skewness and variation in the test residuals. This suggests that the GLM is able to capture some more complex linear relationships that the persistence forecast misses.

\subsection{Neural network (NN)}

A neural network (NN) expands on the GLM by capturing some of the non-linear behaviour highlighted by the large Spearman's correlation coefficients between aggregate V2G plug-in today and aggregate V2G plug-in yesterday in Table \ref{tab:correlation}. Three NNs are trialled: (i) one with the same inputs as the GLM with one-hot encoding applied to the day of the week feature; (ii) one with all of the same features as (i) plus a one-hot encoding of month; and (iii) one with all the same features as (ii) plus a one-hot encoding of hour. The NN with inputs (ii) aims to capture the monthly behaviour displayed in Figure \ref{fig:plug-in_by_month}, while the NN with inputs (iii) aims to capture both the monthly and hourly behaviour displayed by Figure \ref{fig:plug-in_by_month} and Figure \ref{fig:plug-in_by_hour} respectively. However, while the persistence model and GLM developed here are interpretable, the NN-based models are not. This may present challenges when it comes to high-stakes decision making as it is harder to justify the recommendations made by a `black-box' forecast model such as a NN.

\begin{table}[ht!]
    \centering
    \caption{Pearson's and Spearman's correlation coefficients between EV plug-in today and the day before.}
    \begin{tabular}{|c|c|c|}
        \hline
        \textbf{Day} & \textbf{Pearson's} & \textbf{Spearman's} \\
        \hline
        Monday & 0.78 & 0.76 \\
        \hline
        Tuesday & 0.94 & 0.91 \\
        \hline
        Wednesday & 0.95 & 0.91 \\
        \hline
        Thursday & 0.96 & 0.90 \\
        \hline
        Friday & 0.96 & 0.89 \\
        \hline
        Saturday & 0.81 & 0.77 \\
        \hline
        Sunday & 0.84 & 0.78 \\
        \hline
    \end{tabular}
    \label{tab:correlation}
\end{table}

The neural network has two layers with ReLu activation functions; the first layer has 100 neurons, and the second has 50 layers. Each of the hidden layers is followed with a dropout layer dropping 20 \% of units to reduce the risk of overfitting with the small dataset. The loss function is a mean squared error function and the optimizer is an Adam algorithm. The neural network was trained using 1000 epochs and a batch size of 100.

\section{Results \& Discussion}

The root mean squared error (RMSE), mean absolute percentage error (MAPE), and mean absolute error (MAE) displayed in Table \ref{tab:errors} are evaluated to determine the accuracy of the aggregate V2G plug-in forecast. Additionally, the mean, median, standard deviation, range, and interquartile range of the test set residuals displayed in Table \ref{tab:residual_metrics} are used to understand the spread of predictions and the difference between the true values and the predictions.

\begin{table*}[t]
    \centering
    \caption{Root mean squared error (RMSE), mean absolute percentage error (MAPE), and mean absolute error (MAE) for the training, validation, and test sets for each of the persistence model, generalized linear model (GLM), and neural networks.}
    \begin{tabular}{|l|c|c|c|c|c|c|c|c|c|}
    \hline
     \multirow{2}{10em}{\textbf{Model}} &  \multicolumn{3}{|c|}{\textbf{RMSE}} &  \multicolumn{3}{|c|}{\textbf{MAPE  (\%)}} &  \multicolumn{3}{|c|}{\textbf{MAE}} \\
      \cline{2-10}
      & \textbf{Training} & \textbf{Validation} & \textbf{Test} & \textbf{Training} & \textbf{Validation} & \textbf{Test} & \textbf{Training} & \textbf{Validation} & \textbf{Test} \\
      \hline
      \textbf{Persistence} & 10.32 & 10.40 & 10.28 & 10.67 & 10.83 & 10.73 & 7.82 & 8.02 & 7.79 \\
      \hline
      \textbf{GLM} & 8.40 & 8.62 & 8.29 & 8.84 & 8.99 & 8.96 & 6.44 & 6.66 & 6.39 \\
      \hline
      \textbf{NN} & 7.72 & 8.09 & 7.88 & 7.79 & 8.01 & 8.14 & 5.82 & 6.11 & 5.93 \\
      \hline
      \textbf{NN with month} & 5.84 & 6.69 & 6.37 & 5.91 & 6.53 & 6.50 & 4.39 & 5.03 & 4.77 \\
      \hline
      \textbf{NN with month and hour} & \cellcolor{green!25}4.77 & \cellcolor{green!25}6.38 & \cellcolor{green!25}5.96 & \cellcolor{green!25}4.83 & \cellcolor{green!25}6.16 & \cellcolor{green!25}6.07 & \cellcolor{green!25}3.58 & \cellcolor{green!25}4.76 & \cellcolor{green!25}4.46 \\
      \hline
    \end{tabular}
    \label{tab:errors}
\end{table*}

According to Table \ref{tab:errors}, the most accurate model by RMSE, MAPE, and MAE is the NN with month and hour as features, closely followed by the NN with month as a feature but not hour. This suggests that including month as a feature is important for capturing the monthly variations in aggregate V2G plug-in. Contrastingly, including hour as a feature seems less important, especially as there is much more improvement in the training set prediction accuracy compared to the validation and test set prediction accuracies. This suggests that a larger dataset would be beneficial to help avoid overfitting.

\begin{table*}[ht!]
    \centering
    \caption{Mean, median, standard deviation, range, and interquartile range of the test residuals of the forecast models.}
    \begin{tabular}{|l|c|c|c|c|c|}
    \hline
       \textbf{Model} & \textbf{Mean} & \textbf{Median} & \textbf{Standard Deviation} & \textbf{Range} & \textbf{Interquartile Range}\\
    \hline
        \textbf{Persistence} &  -0.24 & \cellcolor{green!25}0.00 & 10.28 & 94.00 & 13.00 \\
    \hline
        \textbf{GLM} & -0.41 & -0.34 & 8.29  & 74.13 & 10.18 \\
    \hline
        \textbf{NN} & -1.75 & -1.54 & 7.68  & 70.44 & 9.05 \\
    \hline 
        \textbf{NN with month} &  0.17 & -0.26 & 6.37 & 53.22 & 7.19 \\
    \hline
        \textbf{NN with month and hour} & \cellcolor{green!25}-0.06 & -0.26 & \cellcolor{green!25}5.96  & \cellcolor{green!25}48.11 & \cellcolor{green!25}6.70  \\
    \hline
    \end{tabular}
    \label{tab:residual_metrics}
\end{table*}

The residual, $r_t$, is calculated using the following equation:
\begin{equation}
    r_t = \hat{y}_t - y_t.
\end{equation}
This means that a negative value is an overestimation and a positive value is an underestimation. Overestimation is concerning as this means that the aggregator may pledge more resources than they will have in actuality resulting in penalty charges. In Table \ref{tab:residual_metrics}, the means and medians of the test residuals are often negative, suggesting that these models all have a tendency to overestimate (apart from the NN with month as a feature which has a positive mean). Probabilistic forecasts may be beneficial to help capture the uncertainty around any forecasts to better inform the aggregator. While the median test residual value is best (i.e. zero) for the persistence forecast, the persistence forecast has a large range of residuals implying that it is often far from the true value. In contrast, the NN with month and hour as features has the lowest standard deviation, range, and interquartile range, and has a mean that is closest to zero. This suggests that the residuals are reduced in magnitude and less variable when a NN with month and hour as features is used to forecast aggregate V2G plug-in.

\section{Conclusion}
\label{sec:conclusion}

This work examined features pertinent to aggregate V2G plug-in and produced a first-of-its-kind day-ahead forecast of minimum aggregate V2G plug-in during 30-minute intervals to aid aggregator participation in ancillary services markets for FR provision. A persistence forecast is developed as a benchmark. This is extended to a GLM which provides a more accurate, yet interpretable, forecast model. Then, the NN-based models further develop this by considering non-linear trends, improving accuracy and consistency at the detriment of interpretability. The NN that includes features based on month and hour provides the most accurate forecasts despite overfitting. 

One pitfall of these models is their tendency to overestimate. A probabilistic forecast that captures the uncertainty on the forecast may give aggregators more information with which to make bids in the ancillary services market. Another solution would be to produce a cost-dependent forecast that takes the penalty fees for failing to deliver promised ancillary services into account. This would curb overestimation by ensuring that the model always aids the interests of the aggregator.

Other avenues that could be explored relating to forecasting aggregate V2G plug-in for FR is extending this model to capture  the anomalous behaviour seen on bank holidays. Additionally, it could be beneficial to forecast what subsection of V2Gs are charging at a given time. Moreover, a regional forecast could help aggregators to provide more targeted FR as would be the case in reality. 



\bibliographystyle{unsrt}
\bibliography{EV_lit_review.bib, PhD_References.bib}

\begin{thebibliography}{10}

\bibitem{Heylen2021}
Evelyn Heylen, Fei Teng, and Goran Strbac.
\newblock Challenges and opportunities of inertia estimation and forecasting in
  low-inertia power systems.
\newblock {\em Renewable and Sustainable Energy Reviews}, 147:111176, sep 2021.

\bibitem{Milano2018}
Federico Milano, Florian Dorfler, Gabriela Hug, David~J. Hill, and Gregor
  Verbic.
\newblock Foundations and challenges of low-inertia systems (invited paper).
\newblock In {\em 2018 Power Systems Computation Conference ({PSCC})}. {IEEE},
  jun 2018.

\bibitem{Chu2020}
Zhongda Chu, Uros Markovic, Gabriela Hug, and Fei Teng.
\newblock Towards optimal system scheduling with synthetic inertia provision
  from wind turbines.
\newblock {\em {IEEE} Transactions on Power Systems}, 35(5):4056--4066, sep
  2020.

\bibitem{NGESO_FFR_website}
{National Grid ESO}.
\newblock {Firm Frequency Response (FFR)}.
\newblock
  https://www.nationalgrideso.com/industry-information/balancing-services/frequency-response-services/firm-frequency-response-ffr?overview.
\newblock Accessed: 01/03/2023.

\bibitem{Kempton2005}
Willett Kempton and Jasna Tomi{\'{c}}.
\newblock Vehicle-to-grid power fundamentals: Calculating capacity and net
  revenue.
\newblock {\em Journal of Power Sources}, 144(1):268--279, jun 2005.

\bibitem{Kempton2005a}
Willett Kempton and Jasna Tomi{\'{c}}.
\newblock Vehicle-to-grid power implementation: From stabilizing the grid to
  supporting large-scale renewable energy.
\newblock {\em Journal of Power Sources}, 144(1):280--294, jun 2005.

\bibitem{Wills2020}
Terri Wills.
\newblock The {UK}'s transition to electric vehicles.
\newblock Technical report, {Climate Change Committee}, 2020.

\bibitem{Meng2015}
Jia Meng, Yunfei Mu, Jianzhong Wu, Hongjie Jia, Qian Dai, and Xiaodan Yu.
\newblock Dynamic frequency response from electric vehicles in the great
  britain power system.
\newblock {\em Journal of Modern Power Systems and Clean Energy},
  3(2):203--211, may 2015.

\bibitem{OMalley2022}
Cormac O'Malley, Luis Badesa, Fei Teng, and Goran Strbac.
\newblock Frequency response from aggregated v2g chargers with uncertain ev
  connections.

\bibitem{OMalley2020}
Cormac O{\textquotesingle}Malley, Marko Aunedi, Fei Teng, and Goran Strbac.
\newblock Value of fleet vehicle to grid in providing transmission system
  operator services.
\newblock In {\em 2020 Fifteenth International Conference on Ecological
  Vehicles and Renewable Energies ({EVER})}. {IEEE}, sep 2020.

\bibitem{Lopes2011}
J~A~P Lopes, F~J Soares, and P~M~R Almeida.
\newblock Integration of electric vehicles in the electric power system.
\newblock {\em Proceedings of the {IEEE}}, 99(1):168--183, jan 2011.

\bibitem{Han2019}
Bing Han, Shaofeng Lu, Fei Xue, and Lin Jiang.
\newblock Day-ahead electric vehicle aggregator bidding strategy using
  stochastic programming in an uncertain reserve market.
\newblock {\em {IET} Generation, Transmission {\&} Distribution},
  13(12):2517--2525, may 2019.

\bibitem{AmaraOuali2022}
Yvenn Amara-Ouali, Yannig Goude, Bachir Hamrouche, and Matthew Bishara.
\newblock A benchmark of electric vehicle load and occupancy models for
  day-ahead forecasting on open charging session data.
\newblock In {\em Proceedings of the Thirteenth {ACM} International Conference
  on Future Energy Systems}. {ACM}, jun 2022.

\bibitem{Dominguez2021}
Donovan Aguilar-Dominguez, Jude Ejeh, Alan~D.F. Dunbar, and Solomon~F. Brown.
\newblock Machine learning approach for electric vehicle availability forecast
  to provide vehicle-to-home services.
\newblock {\em Energy Reports}, 7:71--80, may 2021.

\bibitem{Li2021}
Shuangqi Li, Chenghong Gu, Jianwei Li, Hanxiao Wang, and Qingqing Yang.
\newblock Boosting grid efficiency and resiliency by releasing v2g potentiality
  through a novel rolling prediction-decision framework and deep-{LSTM}
  algorithm.
\newblock {\em {IEEE} Systems Journal}, 15(2):2562--2570, jun 2021.

\bibitem{Perry2021}
Dylan Perry, Ning Wang, and Shen-Shyang Ho.
\newblock Energy demand prediction with optimized clustering-based federated
  learning.
\newblock In {\em 2021 {IEEE} Global Communications Conference ({GLOBECOM})}.
  {IEEE}, dec 2021.

\bibitem{Giordano2020}
Francesco Giordano, Francesco Arrigo, Cesar Diaz-Londono, Filippo Spertino, and
  Fredy Ruiz.
\newblock Forecast-based v2g aggregation model for day-ahead and real-time
  operations.
\newblock In {\em 2020 {IEEE} Power {\&} Energy Society Innovative Smart Grid
  Technologies Conference ({ISGT})}. {IEEE}, feb 2020.

\bibitem{Huber2020}
Julian Huber, David Dann, and Christof Weinhardt.
\newblock Probabilistic forecasts of time and energy flexibility in battery
  electric vehicle charging.
\newblock {\em Applied Energy}, 262:114525, mar 2020.

\bibitem{Jahangir2019}
Hamidreza Jahangir, Hanif Tayarani, Ali Ahmadian, Masoud~Aliakbar Golkar, Jaume
  Miret, Mohammad Tayarani, and H.~Oliver Gao.
\newblock Charging demand of plug-in electric vehicles: Forecasting travel
  behavior based on a novel rough artificial neural network approach.
\newblock {\em Journal of Cleaner Production}, 229:1029--1044, aug 2019.

\bibitem{Saputra2019}
Yuris~Mulya Saputra, Dinh~Thai Hoang, Diep~N. Nguyen, Eryk Dutkiewicz,
  Markus~Dominik Mueck, and Srikathyayani Srikanteswara.
\newblock Energy demand prediction with federated learning for electric vehicle
  networks.
\newblock In {\em 2019 {IEEE} Global Communications Conference ({GLOBECOM})}.
  {IEEE}, dec 2019.

\bibitem{Bessa2013}
R.J. Bessa and M.A. Matos.
\newblock Global against divided optimization for the participation of an {EV}
  aggregator in the day-ahead electricity market. part i: Theory.
\newblock {\em Electric Power Systems Research}, 95:309--318, feb 2013.

\bibitem{Bessa2013a}
R.J. Bessa and M.A. Matos.
\newblock Global against divided optimization for the participation of an {EV}
  aggregator in the day-ahead electricity market. part {II}: Numerical
  analysis.
\newblock {\em Electric Power Systems Research}, 95:319--329, feb 2013.

\bibitem{Clairand2019}
Jean-Michel Clairand.
\newblock Participation of electric vehicle aggregators in ancillary services
  considering users' preferences.
\newblock {\em Sustainability}, 12(1):8, dec 2019.

\bibitem{Clairand2018}
Jean-Michel Clairand, Javier Rodriguez-Garcia, and Carlos Alvarez-Bel.
\newblock Smart charging for electric vehicle aggregators considering users'
  preferences.
\newblock {\em {IEEE} Access}, 6:54624--54635, 2018.

\bibitem{Gautam2019}
Akash Gautam, Arun~Kumar Verma, and Manaswi Srivastava.
\newblock A novel algorithm for scheduling of electric vehicle using adaptive
  load forecasting with vehicle-to-grid integration.
\newblock In {\em 2019 8th International Conference on Power Systems ({ICPS})}.
  {IEEE}, dec 2019.

\end{thebibliography}

\end{document}